\documentclass[preprint]{aastex}

\def\vsini{$v\sin i$}
\def\vsinis{$v\sin i$\space}
\def\oc{$\omega_{crit}$}
\def\ocs{$\omega_{crit}$\space}
\def\m{M$_\odot$}
\def\ms{M$_\odot$\space}

\shorttitle{Angular Momentum in the ONC}
\shortauthors{Tinker, Pinsonneault, \& Terndrup}

\begin{document}
\title{Angular Momentum Evolution of Stars in the Orion Nebula Cluster}
\author{Jeremy Tinker, Marc Pinsonneault, and Donald Terndrup}
\affil{Department of Astronomy, Ohio State University, 
	140 W. 18$^{th}$ Avenue, Columbus, OH 43210, USA}
\affil{E-mail: tinker, pinsono, terndrup@astronomy.ohio-state.edu}

\begin{abstract}

We present theoretical models of stellar angular momentum evolution from the Orion
Nebula Cluster (ONC) to the Pleiades and the Hyades. 
We demonstrate that observations of the Pleiades and Hyades place
tight constraints on the angular momentum loss rate from stellar
winds. The observed periods, masses and
ages of ONC stars in the range 0.2--0.5 M$_\odot$, and the loss
properties inferred from the Pleiades and Hyades stars, are then used to test the initial
conditions for stellar evolution models. We use these models to
estimate the distribution of rotational velocities for the ONC stars at
the age of the Pleiades (120 Myr). The modeled ONC and
observed Pleiades distributions of rotation rates are not consistent if
only stellar winds are included. In order to reconcile the observed loss of angular momentum
between these two clusters, an extrinsic loss mechanism such as
protostar-accretion disk interaction is required. Our model, which evolves the ONC stars with a
mass dependent saturation threshold normalized such that 
$\omega_{crit} = 5.4 \omega_\odot$ at 0.5 \m, and which includes a
distribution of disk lifetimes that is uniform over the range 0--6
Myr, is consistent with the Pleiades. This model for disk-locking
lifetimes is also consistent with inferred disk lifetimes from
the percentage of stars with infrared excesses observed in young
clusters. Different models, using a variety of initial period
distributions and different maximum disk lifetimes, are also compared
to the Pleiades. For disk-locking models that use a uniform
distribution of disk lifetimes over the range 0 to $\tau_{max}$, the acceptable
range of the maximum lifetime is $3.5 < \tau_{max} < 8.5$ Myr. We also
use the ONC period data combined with infrared excess data of the ONC
stars to test this model directly on the pre-main-sequence. At present, the sample of data is
not large enough to make strong conclusions based upon this test.

\end{abstract}


\section{Introduction}

A fundamental problem in the study of stellar evolution is
the loss of angular momentum in stars as they evolve from the
birthline to the main-sequence. Early observations of T Tauri stars
showed rotation rates of $\sim 10\%$ of
their breakup velocity (Vogel \& Kuhi 1981; Bouvier et al. 1986;
Hartmann et al. 1986). New observations of the ONC (Herbst et
al. 2001; Stassun et al. 1999, hereafter SMMV), which increased the
statistical sample of young stars, showed that the majority
of lower mass stars
in this cluster rotate at rates approaching 30\% of breakup. Studies of main sequence
stars (Stauffer \&
Hartmann 1986) and older clusters like the Hyades (Stauffer et
al. 1987; Terndrup et
al. 2000) have shown that late-type stars are mostly slow rotators
(\vsinis $< 20$ km s$^{-1}$). These observations are not
consistent with strict angular momentum conservation as the stars
contract during their evolution 
from the birthline to the main sequence (Stauffer \& Hartmann 1987; Sills, Pinsonneault, \&
Terndrup 2000, and references therein). If the young clusters,
such as the ONC and NGC 2024, represent an earlier evolutionary stage
of older clusters such as the Pleiades and Hyades, then
significant angular momentum loss must occur during the
pre-main-sequence (pre-MS) in order
to produce the slow rotation rates seen at later times. 

It has long been known that stars lose angular momentum through
stellar winds (Schatzman 1962; Weber \& Davis 1967), and this has been used to explain
observed rotation rates at the MS for different mass stars. 
Much theoretical work has been done to explain the angular momentum
loss in stars in the pre-MS (see Krishnamurthi et al. 1997 for a
review). Theoretical models of the interaction
between a rotating central star and a circumstellar accretion disk
have been put forth (K\"onigl 1991; Shu et al. 1994).
According to these models, the star's circumstellar
disk truncates at an inner radius and material is accreted onto the
surface of the star through magnetic field lines. Magnetic torques then
transfer angular momentum away from star and into the disk, causing
the disk and star to corotate at a period that does not change as the
star evolves on the pre-MS. This ``disk-locking'' model naturally produces the slow
rotators seen in observations of young clusters (Collier Cameron \&
Campbell 1993; Keppens et al. 1995).

Observations of the ONC, with an
age of $\sim 1$ Myr, are an important test for these
theoretical models. Attridge \& Herbst (1992) claimed a
distribution of observed rotational periods that was bimodal, with
a group of slow rotators at an average period of $\sim 8$
days, and another group of rapid rotators near 2
days. This bimodality was supported by further studies which increased
the number of measured photometric periods (Choi \& Herbst 1999; Herbst et
al. 2000, hereafter HRHC; Herbst et al. 2001). HRHC used magnetic disk locking to interpret their
distribution of periods, stating that the slow
rotators are stars that are currently locked to their disks, while the
rapid rotators had been released from their disks and were allowed to
spin up. The disk-locking model yields a
rotation period of 8 days for reasonable values of the mass
accretion rate, magnetic field strength and inner truncation radius of
the disk (Shu et al. 1994). 
Their observations are consistent with previous
observations of the distribution of rotational velocities of T Tauri
stars, which showed a peak at slow rotation and a long tail to higher
velocities (Hartmann et al. 1986). The tail in the velocity distribution would lead to a
peak at rapid rotation in a histogram of periods, creating a bimodal
distribution like that seen by HRHC.

Recent observations of the ONC by SMMV have
challenged both the claim of a bimodal period distribution and the
role of stellar accretion disks. Their sample of rotation periods was
statistically consistent with a distribution of periods which is
uniform from $\sim$ 0.5--8 days (see \S2.2 for a brief comparison of
the HRHC and SMMV data). They identified
stars which are actively accreting (and should therefore be
disk-locked) by identifying stars with H$\alpha$ and Ca {\small II} triplet emission in their
spectra. The distribution of periods in this subset of stars was found to be statistically
indistinguishable from the sample of stars without emission. As
another test, near-IR photometry from
Hillenbrand et al.\ (1998) was used to determine which stars had
circumstellar disks. No correlation was found between near-IR 
excess\footnote{i.e. $\Delta$(I-K)$ > 0.3$, where $\Delta$(I-K) is the difference
between the observed I-K value and that expected for a star of a given temperature.}
and rotational period. Since there appeared to be no connection
between the observational signatures of accretion disks and the slow
rotation expected in disk-locking theories, SMMV
challenged the idea that disk-locking is the prime mover of 
angular momentum in pre-MS stars.

Comparisons between theoretical models of stellar rotational
evolution and observations of open clusters (Allian et al. 1996;
Krishnamurthi et al. 1997; Allain 1998;
Barnes, Sofia, \& Pinsonneault 2001) have focused on
solar-mass stars. These studies have demonstrated that there is a
core-envelope decoupling in these stars at an age $\le 10^{8}$ Myr,
and that the saturation threshold of the angular momentum loss from stellar
winds has a dependence on mass. But solar analogues have proved
difficult to model theoretically. In addition to the initial
conditions, the predictions of the models are sensitive to the
treatment of internal angular momentum transport and angular momentum
loss. 

Lower mass stars (M $< 0.5$ \m)  rotate as solid bodies since they
are nearly fully convective up to the time of the Pleiades (120 Myr). Beyond
this age, stars greater than 0.35 M$_\odot$ develop a radiative core,
but most of the moment of inertia is still in the convection zone. 
Solid body models can be
used to constrain the initial conditions of open clusters, such as the
range of disk lifetimes and initial rotation periods, with less
ambiguity. Low-mass open cluster stars exhibit less angular
momentum loss than do high-mass stars, which reduces the influence of
angular momentum loss rates on the inferred initial conditions. Stars below 0.5 \ms show a
wide range of rotational velocities in the Pleiades. 
In the Hyades (600 Myr), rapid rotation persists for low-mass stars,
but those stars between 0.5--1.2
M$_\odot$ rotate at speeds below 10 km s$^{-1}$ (Stauffer et al. 1987;
Terndrup et
al. 2000). Since a range of speeds can still be detected at this age
for low-mass stars, comparisons can be made between a number of
clusters over a wide age span. Thus, we can empirically constrain the
angular momentum loss of pre-MS stars.

In this paper we will present results of models that take the
observed periods for low-mass stars (0.2--0.5 \m) in the ONC 
and evolve these stars forward using theoretical stellar evolution models
and different angular momentum loss rates. In \S2 we present the
theoretical framework of these models and how they were applied to to
the observational data. In \S3 we will apply these models to data from
both the Pleiades and the ONC. These calculations will constrain the
two parameters of the angular momentum loss mechanisms, the saturation
threshold and the disk-locking lifetime, leading to a single preferred model.
Infrared excess data for stars in the ONC is also
used as an additional initial condition to test this model. In \S4 we
discuss the implications of these results, including the uniqueness
of our model and how it can be applied to further observational data
in the pre-MS.

\section{Constructing Theoretical Models}

\subsection{Angular Momentum Loss Mechanisms}

To construct models of low-mass stars, we used the Yale Rotating
Evolution Code (YREC, Guenther et al. 1992). YREC is a Henyey code
which solves the equations of stellar structure in one dimension. YREC
uses the nuclear reaction rates of Gruzinov \& Bahcall (1998) and the
equation of state from Saumon, Chabrier \& van Horn (1995). Our
models have a metallicity of $Z$ = 0.0176 and a mixing length of
$\alpha$ = 1.845, calibrated such that a 1.0 \ms model will
reproduce the solar radius and luminosity at the solar age. The input
physics for these models is discussed in Sills, Pinsonneault, \&
Terndrup (2000).

Although the theoretical models for magnetic star-disk interaction discussed in \S 1 are
sophisticated and complex in their formulation, the treatment of
disk-locking for our angular momentum models is quite simple, since
the disk and star corotate at a fixed angular velocity.
When the star is disk-locked, its period is held
constant over the lifetime of the disk, $\tau_{disk}$, and the angular momentum change
is then a function of star's moment of inertia. When the age of the
stellar model reaches $\tau_{disk}$, the star is released from
disk-locking and evolves under equations (1) and (2) below. The disk
lifetime, $\tau_{disk}$, is relative to the birthline (0 Myr) and not
to the observed age of the star, $\tau_\star$. We use the birthline of
Palla \& Staller (1991), which is the deuterium-burning main sequence
and corresponds to the upper envelope of T Tauri stars in the H-R
diagram. The time evolution of the stellar moment of
inertia was taken from the YREC stellar
models, in which solid-body rotation was enforced. Stellar models of this mass range that included
internal angular momentum transport were nearly identical to solid
body models (Sills, Pinsonneault, \& Terndrup 2000). 

Rotating stars that lose mass through magnetized stellar winds will also
lose angular momentum. To quantify this loss rate, we used a
prescription for angular momentum loss
adopted from Kawaler (1988) and MacGregor \& Brennan (1991), and
described in Krishnamurthi et al. (1997). We write 

\begin{equation}
\frac{d\omega}{dt} = K_\omega \omega^3 \left( \frac{M}{M_\odot}
\right)^{-0.5} \left( \frac{R}{R_\odot} \right)^{0.5} ,\space\space \omega \le
\omega_{crit},
\end{equation}
\
\begin{equation}
\frac{d\omega}{dt} = K_\omega \omega \omega_{crit}^2 \left( \frac{M}{M_\odot}
\right)^{-0.5} \left( \frac{R}{R_\odot} \right)^{0.5} ,\space\space \omega >
\omega_{crit},
\end{equation}

\noindent where $\omega$ is the angular rotation rate of the star, 
$K_\omega$ is a normalization parameter fixed such that a solar
model rotates at the observed solar period at the age of the Sun, and
$\omega_{crit}$ is the threshold at which the angular momentum loss
rate saturates (i.e. $\dot{\omega} \propto \omega$ rather than
$\omega^3$, for $\omega >$ \oc). The constant $K_\omega$ is calculated from Kawaler
(1988) to be $-2.83\times10^{47}$ s.

The value of the saturation threshold for solar mass stars with a
solar convective overturn time has been shown to lie in the range
5--20 $\omega_\odot$ (Patten \& Simon 1996). The value of \ocs scales
as the inverse of the stellar convective overturn time scale (the
Rossby scaling), which is a function of stellar mass. Sills,
Pinsonneault, \& Terndrup (2000) demonstrated that this scaling breaks
down for stars lower than 0.5 \m. These authors empirically measured
the mass dependence of \ocs for low-mass stars by comparing models to
observations of the rapidly rotating stars in open clusters.

Since more data from the Hyades has become available (Terndrup et
al. 2000; Reid \& Mahoney 2000), we will take
only the relative values of \ocs as a function of mass found by Sill,
Pinsonneault, \& Terndrup,  and allow the
zero point of the relation to vary to determine the best values (see
\S3.1). In our presentation of results, we will show models for four
different zero points. To distinguish between these different
normalizations, they will always be
referred to by the value of \ocs at 0.5 \m (see column 1 in Table 1
for these values). For $\omega_{crit}(0.5$\m$) = 3.6$ $\omega_\odot$,
the values of \ocs for 0.4, 0.3, and 0.2 \ms are 2.7, 1.9, and 1.2
$\omega_\odot$ respectively. The numbers for these masses scale linearly with the
values of $\omega_{crit}(0.5$\m$)$ shown in Table 1.

Since the convective time scale of the stars in our
mass range remains nearly constant between the ages of the Pleiades
and Hyades, our assumption that \ocs is a constant for a given mass 
is valid. During the pre-MS phase, convective overturn
time scales are longer, and our
method will tend to overestimate the effect of stellar winds, since
longer convective times will lower \ocs and stars at this stage of
evolution should spend more time in the saturated regime. This
will not significantly affect our results, but places tighter
limits on the ability of stellar winds to reduce angular momentum
in pre-MS stars.

\subsection{Incorporating Observational Data}

The initial conditions for our model are the masses, ages, and periods
of the stars in the ONC, and a distribution of disk-locking lifetimes.
We used the photometric periods listed in HRHC. We chose the periods in this paper
because the stars have been observed over multiple seasons (the total
range of the observations is from 1990--1999), which reduces the
probability of falsely detecting a period to a negligible value. Also,
as discussed by HRHC, some of the data in SMMV
were outside the dynamical boundaries of the ONC, possibly including
stars of different ages than the ONC. The HRHC observations were made continuously within
each observing season (weather permitting), which permits the reliable
detection of periods longer than 8 days. At periods below this value,
there is significant overlap between these data and SMMV.
In the overlap, there is good agreement between the two
sets of data. For comparison, we also considered several analytic
initial period distributions: a flat distribution from 1 to 12 days, a delta
function at 8 days, and a Gaussian curve centered on 8 days with a
standard deviation of 4 days, truncated at 1 and 15 days.

Masses and ages for the HRHC stars were obtained from
Hillenbrand (1997), which were interpolated from the stellar
evolutionary tracks of D'Antona \& Mazzitelli (1994). 
This gave us a sample of 81 stars in the mass
range 0.2--0.5\m. For stars where ages were not available, the mean age of the cluster, 1
Myr, was used. Our calculations are not sensitive to the minor
age spread in the ONC, and running our models with all
stars having an age of 1 Myr did not significantly change the results.

To investigate the effect of disk-locking with our models, we used a
variety of values for $\tau_{disk}$ and for the distribution of of disk
lifetimes, $f(\tau_{disk})$. First, we used a model in which there
were no disks, and models in which all stars had disks with 
the same $\tau_{disk}$. Then we implemented a model for 
$f(\tau_{disk})$ motivated by recent observations of young clusters.
Haisch, Lada \& Lada (2001) reported results of JHKL
photometry of clusters ranging in mean age from $\sim$ 0.5--5 Myr. They
examined the fraction of stars in each cluster with the infrared
excess indicative of cicrumstellar disks. The initial disk fraction is
very high ($\ge$ 80\%) and decreases linearly with age to an
extrapolated maximum disk lifetime of $\tau_{max} \sim$ 6 Myr. Their results imply that
all stars are born with accretion disks around them and that the
spectrum of disk lifetimes is flat in this range of 0--6 Myr.

With the initial conditions of mass, age, and initial period, the
rotation of an observed star can be computed at any future time for a given
\ocs and $\tau_{disk}$.
At the age of the Pleiades, the projected period of the ONC star is
converted to an equatorial rotation velocity and then multiplied by a
random $\sin i$, generated by producing a random number
$0<p<1$ and calculating $\sin i=(2p-p^2)^{1/2}$. This allows for
statistical comparison between the modeled ONC and the observed
distribution of \vsinis from the Pleiades taken from Terndrup et al. (2000).

To test the hypothesis that the projected distributions
of \vsinis created by our models are inconsistent with an observed
\vsinis distribution, two-distribution K-S tests were used. We defined limits of
P$_{KS} < 0.01$ (99\% confidence level) and P$_{KS} < 0.10$ (90\%
confidence level) for rejecting the hypothesis that the two distributions
are taken from the same parent distribution. These limits are arbitrary,
but the K-S test was used extensively in our results and these limits
add consistency and show how our results are constrained at different
confidence levels.

\section{Evolving Open Clusters Forward in Time}

The two angular momentum loss mechanisms described in \S2.1 affect stars
on different time scales. Therefore we applied our technique to 
systems at very different ages.
To evaluate the stellar wind angular momentum loss rates without the use of
disk-locking, we computed the expected distribution of rotation rates
for the Pleiades cluster at 600 Myr. In these
models, presented in \S3.1, the value of \ocs was varied and the
resulting rotational distributions were compared to the Hyades data.
We then constructed models projecting the ONC forward to the age of
the Pleiades. These models, presented in \S3.2, varied both \ocs and
$\tau_{disk}$. We present the results of models that take
$\tau_{disk}$ as a constant and then employ the distribution of
disk lifetimes motivated by observations.

The models presented in \S3.2 make no prior assumptions about the angular
momentum loss of a star in the ONC between the birthline and the
current epoch. We did not assume that the slow rotators are disk-locked
and the rapid rotators are free of their disks. All stars of any
period are equally likely to have an accretion disk of a given
lifetime. We argue that this is consistent with the observations of SMMV but with a
very different interpretation than theirs.
To check our results against this effect, \S3.3 presents a model of the ONC
which used the additional initial condition of the infrared data of
Hillenbrand et al. (1998) to infer which ONC stars can be allowed to
have disks. In \S3.4 we will present
models that start at the birthline and use various initial
period distributions, comparing these models both to the ONC and the Pleiades.

\subsection{Projecting the Pleiades to the Hyades}

To model the evolution of rotation rates in the Pleiades,
the same mass range of 0.2--0.5 M$_\odot$ was used, as well 
as the same procedure as described above for the
ONC stars, but using \vsinis as an initial condition rather than
rotation period. Since the age of the Pleiades is much greater
than the time scale for disk lifetimes, disks will
have little effect on the evolution of these stars.
These calculations place strong constraints on the
saturation threshold, $\omega_{crit}$. Since both the Pleiades and
Hyades data are subject to inclination effects, projecting the Pleiades stars forward gives an
expected distribution of \vsini, which is what we have to compare
from the Hyades (also taken from Terndrup et al. 2000\footnote{ This calculation is similar to the one described \S 4 of
Terndrup et al. (2000) for stars less massive than 0.4 M$_\odot$. But
it differs from the calculation in this paper in that homologous
contraction was assumed for all stars as they evolved. It was also
assumed that all stars were in the saturated regime during their
evolution. }).
The detection limit for the Hyades observations was $v\sin i = 6$ km s$^{-1}$,
so this number was used as a minimum value for the evolved Pleiades
stars. Figure 1 shows the expected
\vsinis distribution of the Pleiades for four values of
\oc. Values of \ocs $= 1.8, 3.6$, and $ 7.2
\omega_\odot$ are excluded by K-S tests at 99\% confidence, while
\ocs $= 5.4 \omega_\odot$ has a K-S probability above both limits. At
the 90\% confidence level, the allowed range of saturation thresholds
is $4.5 \omega_\odot < \omega_{crit} < 6.5 \omega_\odot$ at 0.5 \m. We
adopted 5.4 $\omega_\odot$ as our best value, but for the models of the
ONC, we still considered a range of values to test the
sensitivity of these calculations to the choice of \oc.

\subsection{Projecting the ONC to the Pleiades}

The observations of circumstellar disks suggest that these disks
do not typically exist far beyond ages of $\sim$ 6 Myr. For our initial models
of ONC, we chose values of $\tau_{disk}$ of 0, 3, and 6 Myr. The choice
of $\tau_{disk} = 0$ Myr is motivated to test the idea that stellar
winds alone can account for the angular momentum loss between 1 and
120 Myr. Disks of 6 Myr and 3 Myr were chosen to explore the maximally allowed
and intermediate scenarios respectively. The four values of \ocs
used in \S3.1 were also used here. The combination of these parameters
gives us 12 different angular momentum loss rates.
Figure 2 shows the cumulative distribution of \vsinis for the low-mass ONC
stars projected forward to an age of 120 Myr for these models. 
Figure 2 is a grid of different models,
with the rows being saturation thresholds of 1.8, 3.6, 5.4, and 7.2
$\omega_\odot$, and the columns being disk locking lifetimes of 0, 3,
and 6 Myr.

\begin{deluxetable}{cccc}
\tablecolumns{4} 
\tablewidth{0pc} 
\tablecaption{K-S results from Figure 2} 
\tablehead{ 
\colhead{$\omega_{crit}(0.5$M$_\odot)$} & \colhead{$\tau_{disk}$ = 0 Myr} & \colhead{$\tau_{disk}$ =
3 Myr } & \colhead{$\tau_{disk}$ = 6 Myr}
}
\startdata
1.8 $\omega_\odot$  & $1.5\times 10^{-7}$ & 0.31 & $2.7\times 10^{-3}$ \\
3.6 $\omega_\odot$ & $1.3\times 10^{-9}$ & 0.97 & $2.7\times 10^{-3}$ \\
5.4 $\omega_\odot$ & $6.2\times 10^{-8}$ & 0.11 & $1.7\times 10^{-4}$ \\
7.2 $\omega_\odot$ & $9.4\times 10^{-6}$ & 0.11 & $6.3\times 10^{-6}$ \\

\enddata
\end{deluxetable}

Table 1 shows the values returned by the K-S test for all twelve
panels in Figure 2. The distributions in the first column, 
showing the projections of the
ONC at an age of 120 Myr with no disk locking, are not compatible with the observed \vsinis
distribution of the Pleiades at 99\% confidence. Indeed, a significant fraction
of the stars would be rotating at or near breakup velocity.
The projections shown in the panels in the right-hand column,
where all stars had disks that
lasted for 6 Myr, are also excluded at the same confidence. The
results of the models for 3 Myr disks are quite good, with K-S values
above the 0.1 level for all four values of \oc.
However, to assume that all stars are locked
to disks of the same lifetime is too simplistic to be taken as a
realistic model, and it does not agree with independent direct
estimates of accretion disk lifetimes. But Figure 2
clearly shows that angular momentum loss through magnetic stellar
winds alone cannot account for the observations in young stellar
clusters, even when the entire range of possible values for \ocs is
explored. An additional loss mechanism is required.

To implement the distribution of disk lifetimes discussed in \S2.2, 
we created Monte Carlo simulations of the
expected \vsinis distribution of the ONC at 120 Myr in which
$\tau_{disk}$ could vary for each star. In these
simulations each ONC star was projected forward in time with an \ocs
of 5.4 $\omega_\odot$, a randomly sampled disk lifetime in the range
0 to $\tau_{max}$, and a randomly generated $\sin i$. For a model where
$\tau_{max} = 6$ Myr, we performed 1000 simulations and compared the average
distribution to the \vsinis distribution of the
Pleiades. These two distributions are plotted in Figure 3. For comparison,
Figure 4 plots \vsinis versus mass for a sample
simulation from the 1000 that were generated, as well as for a model
which contained no disk-locking. This figure also
includes a plot of \vsinis versus mass for the Pleiades
observations.

The maximum age to which accretion
disks can live and also still govern the rotation of the star is
an important and unresolved question. We ran a series of Monte Carlo simulations in
which the maximum disk lifetime varied from 0 Myr (no disks) to a
distribution that reaches to 12 Myr. Figure 5
shows the results of the simulations. For each maximum disk lifetime,
we ran 1000 Monte Carlo simulations and compared the average \vsinis
distribution to the Pleiades with two-distribution K-S
tests. As can be seen in Figure 5, distributions of disks in which
the maximum lifetime is below 3 Myr are excluded at 99\% confidence, 
as well as maximum disk lifetimes above 9 Myr. The region of high K-S probability centers at 6.5
Myr, which is in good agreement with the observations. The range of
maximum disk lifetimes above the 90\% rejection level is $3.5$ Myr $< \tau_{max} < 8.5$ Myr.  

\subsection{Using Individual Stellar IR Data}

It is important to determine if our model of the ONC is consistent when
adopting the added constraint of the individual infrared measurements
for each star used in our sample. We
incorporated the IR excess data from
Hillenbrand et al. (1998) that was used in both HRHC and SMMV to
distinguish between stars with and without disks. The ONC was once again projected to the
time of the Pleiades, using the infrared data as an additional input
parameter. Stars with IR excess $\Delta$(I-K)$ > 0.3$ were allowed to
have disks with lifetimes in the range
$\tau_\star$ to $\tau_{max}$. Stars with $\Delta$(I-K) below this
threshold were evolved with stellar winds only. For the stars with no IR data
(10\% of the sample) disks were randomly sampled from the range
0 to $\tau_{max}$. For 81 the stars used in this
study, 70\% had excess IR emission. The resulting expected ONC
distribution for $\tau_{max} = 6$ Myr was consistent with the Pleiades
by K-S test.

For our sample of ONC stars, the two period distributions of stars above
and below the IR limit are statistically indistinguishable by K-S test,
although the mean period is slightly smaller
for stars below the IR limit. Figure 6 shows the cumulative
distributions for stars above and below the IR excess limit. The similarity of these two
distributions shows that our prescription of assigning disks to stars
regardless of IR data did not bias our results.

If our assumption is true that all stars are born while locked to disks, and if
the diagnostic $\Delta$(I-K) is a true indicator of disks, then the
distribution of periods for stars with $\Delta$(I-K)$ > 0.3$ is a true
sample of rotation periods at the birthline (since they are still
locked to their disks, their periods are the same as they were at 0
Myr). As another test of our model, we used these stars above
the IR limit as a birthline distribution and projected it forward to
the time or Orion using a uniform distribution of disk lifetimes
between 0 and 1 Myr.
This projected distribution should be consistent with the ONC stars
below the IR limit, since at the birthline they too had disks, but
were released from them before the current age of the ONC. A K-S test
shows these two distributions to be indistinguishable as well. 
If this same calculation is performed without disk-locking, however,
distinct differences are seen between the expected period distribution
and the observed distribution of stars below the IR limit. Even with
the small statistics, a K-S test rejects the consistency of these
distributions at 99\% confidence. The cumulative period distributions
calculated with these two models are also plotted in Figure 6. The
small sample size inhibits any strong conclusions to be made from
these tests; only 17 stars in our sample were below the IR excess limit.
But the radical difference between the two models in Figure 6, and the
difference between the no-disk model and the 17 stars below the IR
limit, is apparent.

Though readily available, the use of $\Delta$(I-K)$ > 0.3$ as a proxy
for circumstellar disks is not as strong a criterion as other techniques.
The observations of Haisch, Lada \& Lada (2001)
used JHKL-band photometry. The addition of the L-band in their
analysis makes their results a more robust indicator of
stars with circumstellar disks (Haisch, Lada \&
Lada 2000). Therefore we believe that future observational tests should
utilize L-band data in order to provide a cleaner test for the
presence of a massive disk.

\subsection{Rotation of Stars at the Birthline}

If stars at the birthline were all locked to accretion disks rotating
at the canonical disk-locking period of 8 days, it would be difficult to explain
the observations of either the ONC or the Pleiades together with the
observed distribution of disk lifetimes. For a distribution of disk
lifetimes ranging from 0--6 Myr, less than 1/6 of the stars in the ONC
would have been released from their disks and had the opportunity to
spin up, not enough stars to populate the rapid rotator portion of the
HRHC data. We also compared what a population of
stars, all disk-locked with an initial period of 8 days, would look
like at the time of the Pleiades using the same techniques described
in \S2.2. (We used the same sample of stars from the ONC as before, this
time using the mass of the star only and assuming an age of 0 Myr.)
The results are plotted in Figure 5, showing P$_{KS}$ versus maximum
disk lifetime. The curve peaks sharply at 4.5 Myr and has a narrow
range of inclusion. For comparison, the same calculation was performed for
an initial distribution of periods that was uniform
in the range 1--12 days. It is plotted on Figure 5 as well. At the time
of the Pleiades, an initially flat distribution of periods is compatible with a
maximum disk lifetime of 6 Myr. Maximum disk lifetimes below 3 Myr are
excluded just as with the ONC data, but a flat period distribution
does not place strict upper limits on $\tau_{max}$. The presence of
rapid rotators at the birthline creates a need for long-lifetime
disks. The period distribution presented in SMMV, which was shown to be
consistent with a uniform distribution, would also require longer
disk-locking lifetimes.

A uniform period distribution and a delta function at 8 days are
extreme possibilities for rotation periods at the birthline.
Even though a value of 8 days is preferred by disk locking theories, 
the variables which are used to calculate the rotation of the
accretion disk may vary, and it is more realistic to assume
that the initial distribution of periods would be a
Gaussian. A distribution of this form centered on 8 days with a
standard deviation of 4 days (cut off below 1 day and above 15 days) 
was evolved from the birthline to the
Pleiades, once again using the masses of the ONC stars. 
The results, plotted on Figure 5, strongly exclude low and high values
for the maximum disk lifetime, with an area of inclusion centered on
5.5 Myr.
It is possible to compare the period distribution of the ONC
to the Gaussian birthline distribution as well. Figure 7 shows
this comparison. The cumulative distributions, shown in the lower panel,
have a P$_{KS} = 0.54$. Our choice of 8 days as the mean period was
obviously motivated by theoretical preferences and to provide a good
fit to the HRHC data. It is possible to lower this mean down to 6.5
days and still achieve similar results.

\section{Discussion}

It is an inherent
assumption throughout this paper that when the ONC reaches an age of
120 Myr it {\it should} look like the Pleiades as seen today,
i.e. that they have the same initial conditions, disk lifetimes, and
wind properties. There are uncertainties in
the comparison of two different clusters; the mass functions could
differ significantly, different formation mechanisms or environments
could lead to different evolutionary patterns, and the observations of
each cluster may have systematic differences. But by restricting our
calculations to a narrow range in mass in which stars rotate as solid
bodies and evolve through the pre-MS phase in similar time scales,
the comparison is valid and useful.

The models of the later-age Pleiades cluster are very effective at
placing limits on the saturation threshold, but say little about
disk-locking lifetimes. The models of the ONC are relatively
indifferent to the choice of \oc, producing rotational distributions
that are similar for models where nearly all stars are in the
saturated regime the majority of the time (1.8 $\omega_\odot$), and
models in which only the rapid rotators become saturated (7.2
$\omega_\odot$). Instead, for the ONC models, the dominant variable is
the disk-locking lifetime, even though the maximum allowed
$\tau_{disk}$ is much smaller than the age of the Pleiades. 

Our results show that it is difficult, if not impossible, to reconcile
the observed amount of angular momentum loss between the ages of the
ONC and the Pleiades without the use of a loss mechanism in addition to
stellar magnetic winds. This result is similar to that found by Rebull
et al (2001), in which they analyzed ONC stars of a larger mass
range. In \S2.1
we stated that our prescription for \ocs would tend to overestimate the
effects stellar winds during the pre-MS. Since the models of the ONC presented in the first column
of Figure 2 all fail to match the Pleiades, even with this
overestimated loss rate, the conclusion that stellar winds alone cannot
explain the observed rotation rates is strengthened.

If the additional loss mechanism is disk locking,
it is also difficult to reconcile the observations if the disk
lifetimes are always less than 3 Myr or can be greater than 8 Myr. The
intersection of the constraints made by the ONC and Pleiades models
give us the parameters of \ocs$= 5.4 \omega_\odot$ and $\tau_{max} = 6$
Myr for our preferred model. This model, however, is not a unique
solution to the problem. For example, the column of models in Figure 2
where $\tau_{disk}$ = 3 Myr for all stars could not be excluded by K-S
test at our defined limits. It is strictly an observational constraint
on $\tau_{disk}$ from IR data that this model is considered invalid. 

Different distributions with other functional
forms of $f(\tau_{disk})$ or $\tau_{max}$, combined with
different initial conditions such as those presented in \S4.3, can be
constructed to compare to the data as well. Further constraints on
disk lifetimes and the initial period distribution can only be made
through larger statistical samples. The number ONC stars with
photometric periods will soon be increased by $\sim$ 400 (Herbst et
al. 2001). With this expanded data set, tests like those in \S3.4 will become a powerful
tool in constraining angular momentum loss through disk-locking.  

One of the reasons for the recent controversy
over the observations of the ONC, namely the bimodal period
distribution, is not evident in the subset of the HRHC data used in
this paper. As shown in Figure 7, the ONC periods used were consistent
with the Gaussian birthline model presented in \S3.4. At our defined
levels of acceptance, the K-S test does
not rule out the hypothesis that the observational distribution was
drawn from a uniform distribution as well.

Our preferred model does not predict that a bimodal distribution would be
apparent at 1 Myr for this range of masses. The results reported in
Herbst et al. (2001) shows that the mean period of stars in the ONC is
mass dependent. The bimodal period distribution could then be an
effect of plotting a wide range of masses in one histogram. The high
mass stars have different evolutionary time scales, internal angular
momentum transport, and possibly different disk-locking lifetimes or
initial conditions. The aforementioned increase in the ONC period
sample would allow smaller ranges of mass to be grouped together in
order to test these effects.

Our model does predict that the rotation rates of a cluster in
the later stages of the distribution of disk lifetimes, such as NGC
2264 and NGC 2362 (3.2 and 5.0 Myr respectively), would show a
bimodality caused by the spin-up of stars released from their
disks. Photometric period observations of clusters such as these would
be key to further constraints on disk-locking lifetimes. Another
important test would be to incorporate observations of a cluster just
beyond the observed maximum disk lifetime. Rotational measurements of
stars at this age would constrain $\tau_{disk}$ by comparing models of
younger clusters using the method presented here. Observations at this
age would place further constraints on loss rates from stellar winds
since this mechanism would have to account for all the angular momentum
loss from that point to the Pleiades and Hyades.

\acknowledgements This work was supported by NSF grant
AST-9371621 (M. P. and D. T.). D. T. would like to thank Robert Mathieu and Keivan
Stassun for helpful discussions.

%===================================================================

\begin{figure}
\plotone{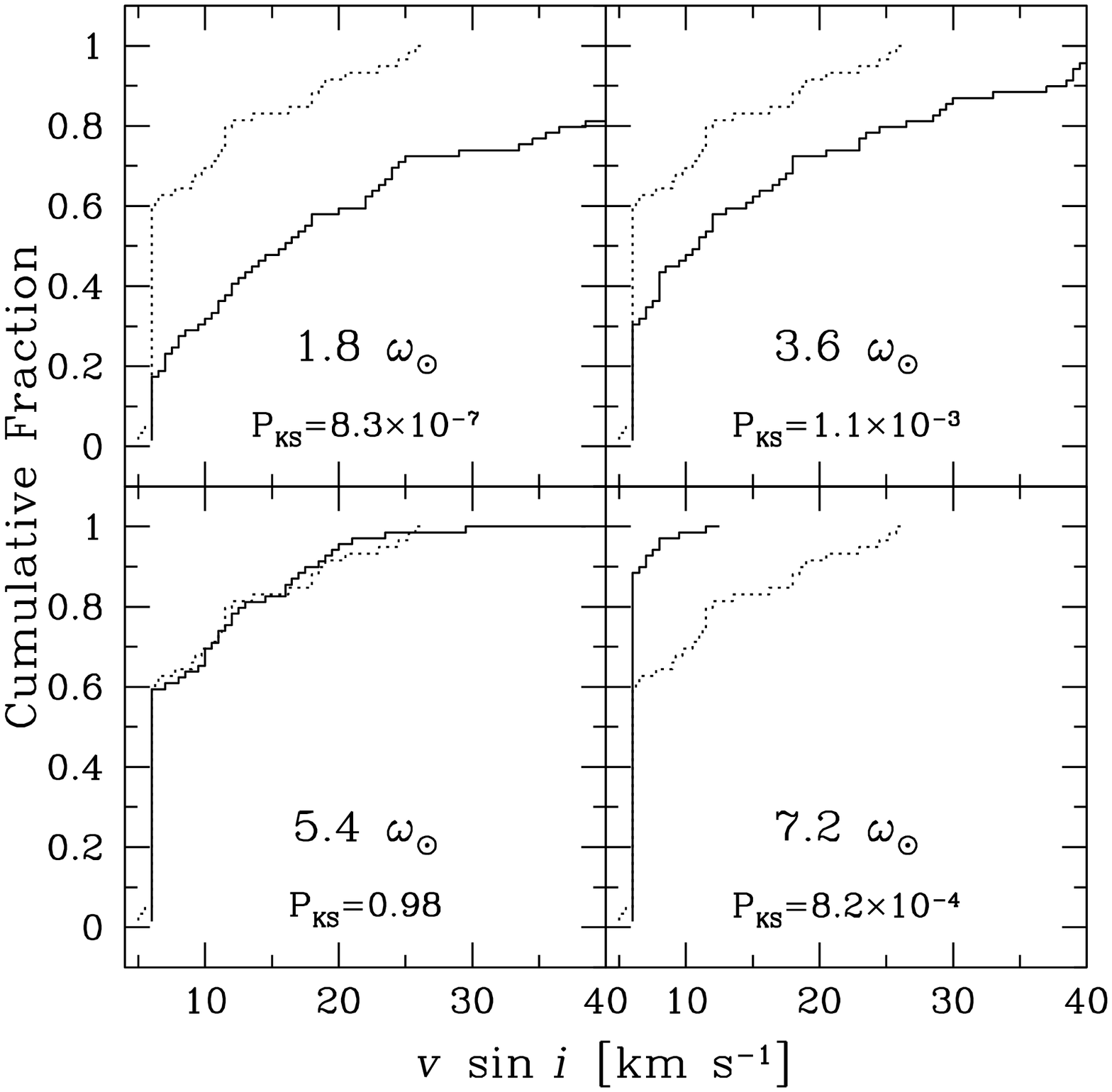}
\caption{\small The cumulative distribution of \vsinis for the projected
sample of Pleiades stars is shown for 4 different saturation thresholds. The projected
Pleiades is the solid line and the observed Hyades distribution is the
dotted line.}
\end{figure}

\begin{figure}
\plotone{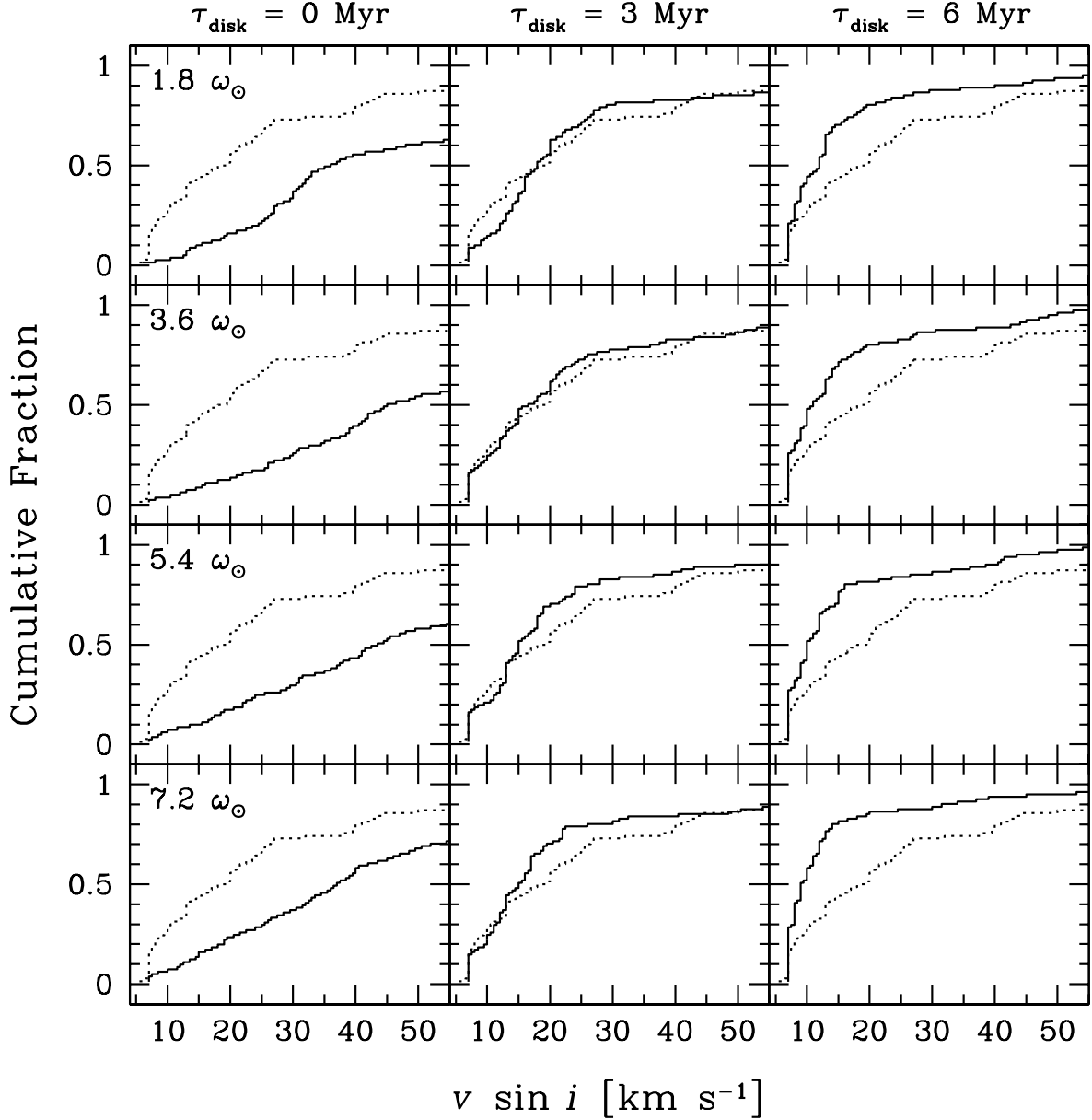}
\caption{\small The cumulative distribution of \vsinis for the projected
sample of ONC stars is shown for 12 different loss laws. The projected
ONC is the solid line and the observed Pleiades distribution is the
dotted line. The figure is a grid of models, with the columns being
different disk-locking lifetimes and the rows different saturation thresholds.}
\end{figure}

\begin{figure}
\plotone{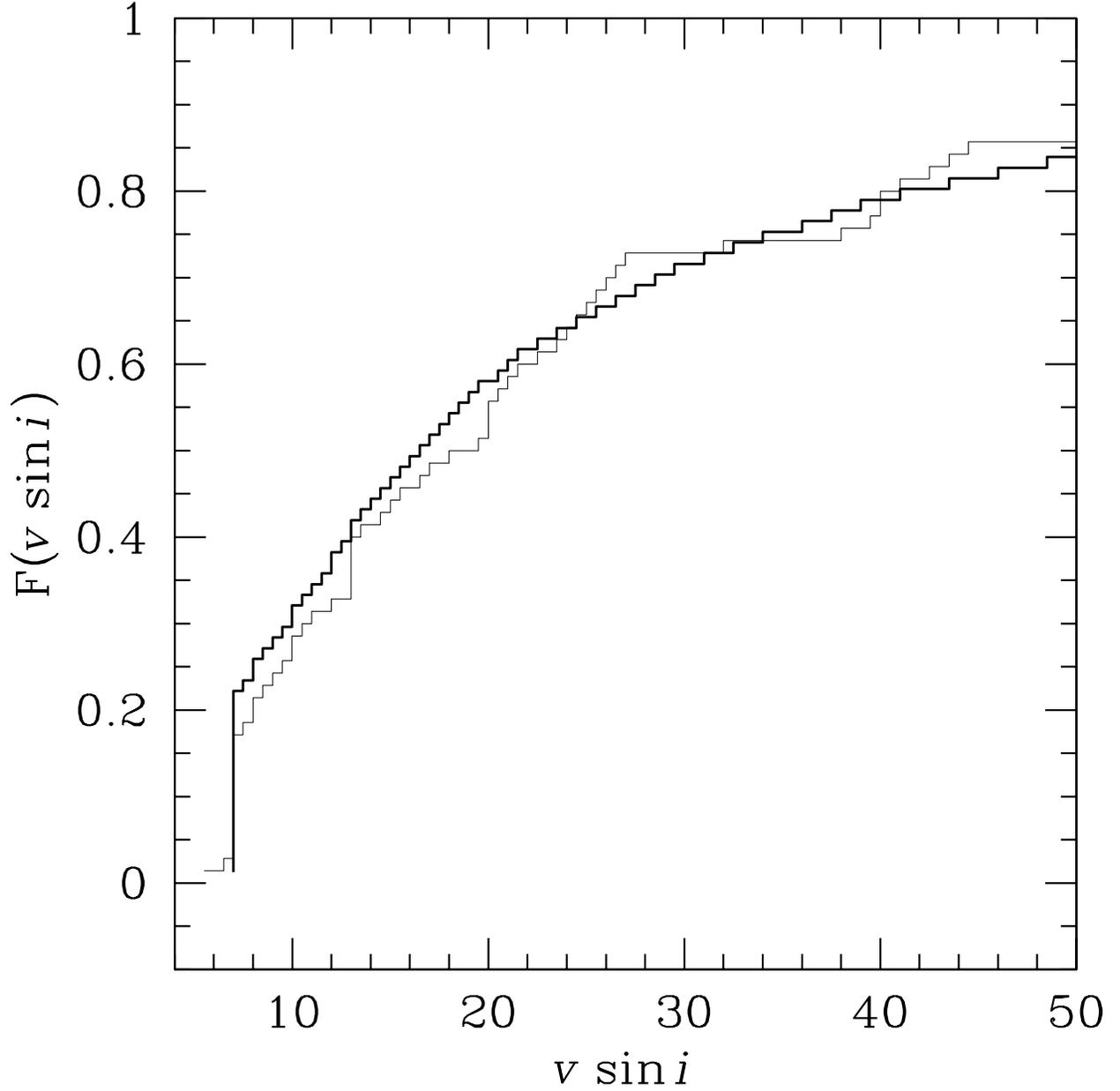}
\caption{\small The cumulative distribution of \vsinis for the
projected ONC is shown with the dark line. This curve is the average
distribution for 1000 Monte Carlo simulations. The model used $\omega_{crit} =
5.4 \omega_\odot$ and a uniform distribution of disk lifetimes of 0--6
Myr. The thin line is the observed distribution of \vsinis for the
Pleiades. The P$_{KS}$ for these two distributions is $0.86$.}
\end{figure}

\begin{figure}
\plotone{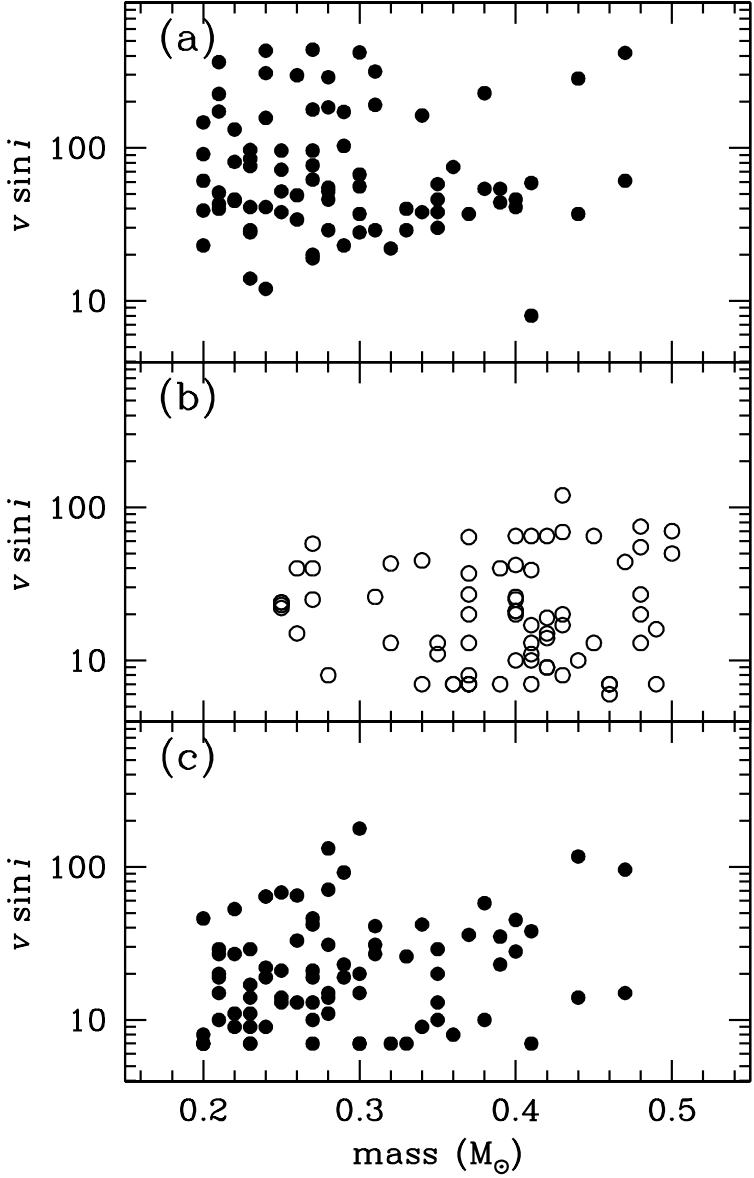}
\caption{\small (a) \vsinis
is plotted against mass for a sample Monte Carlo simulation projecting
the ONC forward to the time of the Pleiades. This model used no
disk-locking. Masses are taken from Hillenbrand (1997). (b) The
observed values of \vsinis and mass for the Pleiades, taken from
Terndrup et al. (2000). (c) \vsinis versus mass for a sample
simulation taken from the average distribution plotted in Figure
3. This model used a uniform distribution of disk lifetimes ranging
from 0--6 Myr.}
\end{figure}

\begin{figure}
\plotone{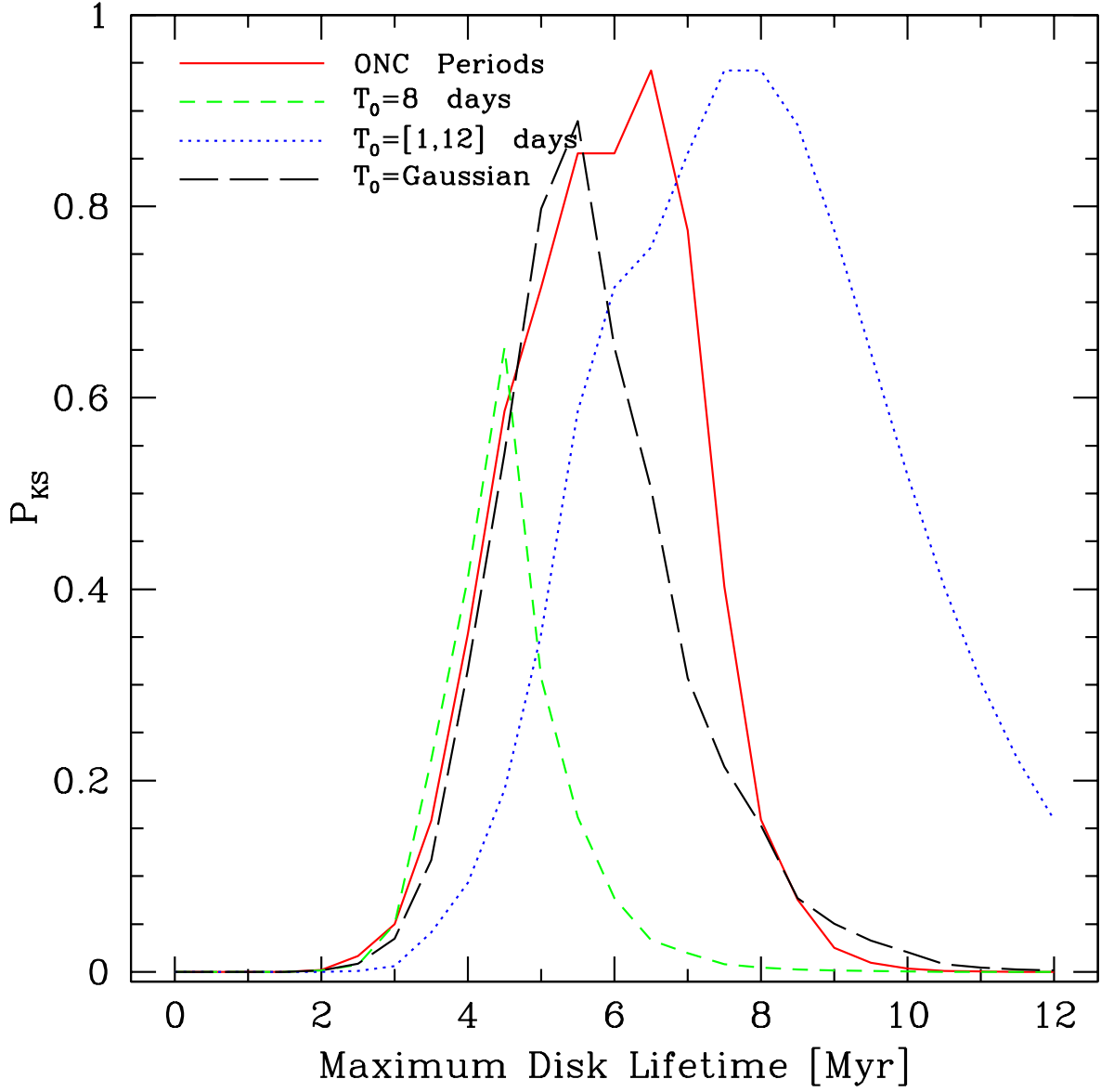}
\caption{\small The K-S probability for the evolved simulations and observed
Pleiades is plotted as a function of maximum disk lifetime (for
distributions of disks of the range 0 to $\tau_{max}$). The solid line
is the evolved ONC stars. The dotted line is for a model started at
the birthline with a initially uniform distribution of periods in the
range 1-12 days. The short dashed line is for a model started at the
birthline with all starting periods being 8 days. The long dashed line
is a birthline model using a Gaussian distribution of initial
periods, which can be seen in Figure 5. The masses of the stars for
all birthline models are the same as the ONC stars.}
\end{figure}

\begin{figure}
\plotone{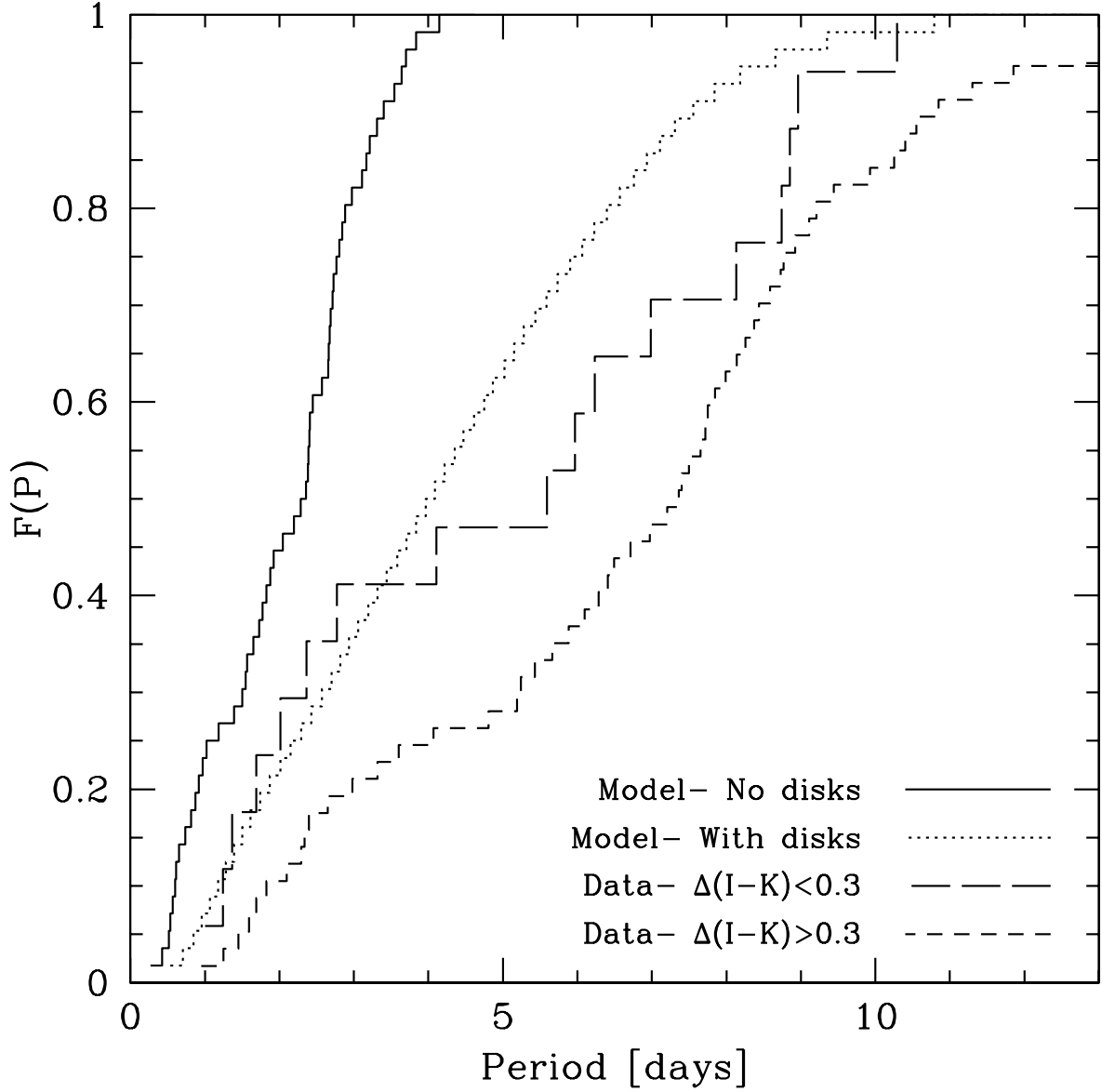}
\caption{\small The cumulative distributions of periods from HRHC for
stars with IR excess ({\it short-dash line}) and without IR excess
({\it long-dash line}) are plotted. The projected distribution of
periods at the age of Orion for two models, one with no disk-locking
({\it solid line}), and one with disk lifetimes of 0--1 Myr ({\it
dotted line}) are shown as well. The models used the periods of the 57 stars with IR
excess as a birthline distribution. }
\end{figure}

\begin{figure}
\plotone{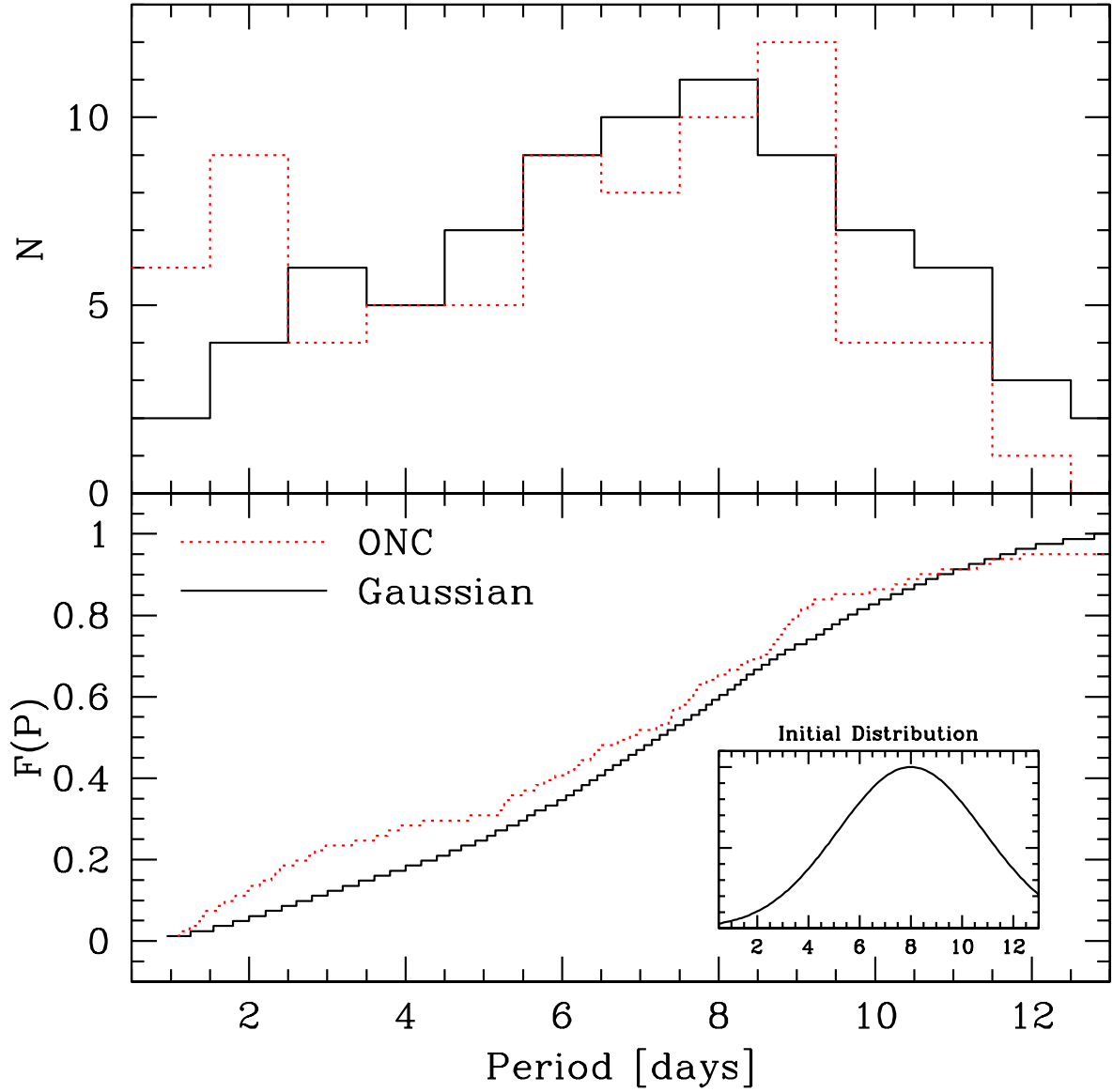}
\caption{\small In the upper panel, the observed period distribution from HRHC for
low-mass stars is plotted ({\it dotted line}) along with the period distribution for a
model started at the birthline ({\it solid line}) using a Gaussian initial period
distribution (shown in the inset box in the lower panel). The
cumulative distributions are plotted together in the lower panel. The
P$_{KS}$ for these two distributions is 0.54.}
\end{figure}

\end{document}